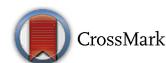

# GPU-accelerated Gibbs sampling: a case study of the Horseshoe Probit model

Alexander Terenin[1] · Shawfeng Dong[2] · David Draper[2]

**Abstract**
Gibbs sampling is a widely used Markov chain Monte Carlo (MCMC) method for numerically approximating integrals of interest in Bayesian statistics and other mathematical sciences. Many implementations of MCMC methods do not extend easily to parallel computing environments, as their inherently sequential nature incurs a large synchronization cost. In the case study illustrated by this paper, we show how to do Gibbs sampling in a fully data-parallel manner on a graphics processing unit, for a large class of exchangeable models that admit latent variable representations. Our approach takes a systems perspective, with emphasis placed on efficient use of compute hardware. We demonstrate our method on a Horseshoe Probit regression model and find that our implementation scales effectively to thousands of predictors and millions of data points simultaneously.

## 1 Introduction

The Bayesian statistical paradigm has a variety of desirable properties. It accounts for the uncertainty inherent in statistical inference by producing a posterior distribution, which fundamentally contains more information about the unknown quantities of interest than a point estimate. It also propagates this uncertainty to predictive distributions and thus does not overfit in a way that paradigms that produce only point estimates may do. Unfortunately, the computational methods required to produce a posterior distribution tend to be expensive. In particular, Markov chain Monte Carlo (MCMC) methods (Metropolis et al. 1953; Hastings 1970; Geman and Geman 1984)—the cornerstone of modern Bayesian computation—often do not scale well either with data set size or with model complexity.

In this paper, we present a case study of a way to implement MCMC for a large class of Bayesian models that admit exchangeable likelihoods with latent variable representations. We do so by performing the computations on a graphics processing unit (GPU), a widely available parallel processor originally designed for 3D video use cases but well suited to a variety of other tasks. In the sections that follow, we describe GPUs, characterize models in which this approach is usable, and demonstrate the method on a Horseshoe Probit model with $N = 1{,}000{,}000$ and $p = 1000$. Standard computation with such $N$ may easily take $O$ (days)—the method we describe runs in $O$ (minutes).

This approach requires no new mathematical theory—instead, we consider the *systems perspective*. MCMC is widely thought to be inherently sequential and unsuited to parallel environments. Furthermore, many practitioners are not aware that substantially different approaches can be needed to parallelize algorithms for use on GPUs, rather than computer clusters, due to issues such as *warp divergence* that will be described later. To compare with an approach well-suited to compute clusters, see Terenin et al. (2016). Our contribution here is to demonstrate that Gibbs sampling on GPUs is doable for a generic class of models and to present ideas that a practitioner would need to consider in imple-

✉ Alexander Terenin
a.terenin17@imperial.ac.uk

Shawfeng Dong
shaw@ucsc.edu

David Draper
draper@ucsc.edu

[1] Statistics Section, Department of Mathematics, Imperial College London, London, UK
[2] Applied Mathematics and Statistics, University of California, Santa Cruz, Santa Cruz, CA, USA

menting GPU Gibbs sampling, in the context of a Horseshoe Probit regression model.

## 2 Previous work

There are a number of approaches for using GPUs for accelerating computation in Bayesian inference—some approaches are completely model specific, others are generic. The following overview is brief, emphatically not exhaustive, and presented in no particular order.

- *Bayesian mixture models.* Suchard et al. (2010) review and describe GPUs, and outline a method for performing calculations needed to fit Bayesian mixture models with MCMC on a GPU.
- *Hamiltonian Monte Carlo.* Beam et al. (2015) outline a method for fitting a Bayesian multinomial logistic regression model on a GPU with Hamiltonian Monte Carlo.
- *Parallel tempering.* Mingas and Bouganis (2012) describe a method for sampling from multimodal distributions with parallel tempering, using hardware acceleration in the form of a field-programmable grid array (FPGA), and compare their method with GPU implementations.
- *Sequential Monte Carlo.* Lee et al. (2010) review the architecture, programming model, and performance of GPUs and describe methods for running importance-sampling-based algorithms such as sequential Monte Carlo on GPUs.
- *State-augmented marginal estimation.* Seita et al. (2015) propose a way to use GPUs to accelerate an annealed variant of Gibbs sampling, used for obtaining high-quality point estimates without getting stuck in local optima in discrete state spaces.
- *Latent Dirichlet allocation (LDA).* A number of authors have created GPU-accelerated algorithms for LDA. These include Yan et al. (2009) (using Collapsed Gibbs sampling) and Canny and Zhao (2013) (using variational Bayes).
- *Deep learning.* Krizhevsky (2014) outlines a way to employ GPUs to accelerate algorithms used for deep learning. Though these algorithms are not Bayesian, in the computational sense of producing a distribution function as output, they are meaningful to include because the success of deep learning illustrates the sheer complexity of models that can be fitted with GPUs.

## 3 Review and description of GPUs

A GPU can be thought of as a massively parallel coprocessor. Whereas a CPU typically has a few physical cores optimized for sequential serial processing, a modern GPU has thousands of smaller cores optimized for handling multiple tasks simultaneously. GPUs operate according to the *single-instruction-multiple-thread (SIMT)* (Matloff 2011) model, a special case of *single-instruction-multiple-data (SIMD)* (Matloff 2011): every thread in a thread block must execute the same instructions, but with respect to a different location in memory. Multiple thread blocks execute concurrently on the same GPU. Whereas a CPU is designed to quickly perform a sequence of instructions with low latency, a GPU is designed to perform data-parallel, high-throughput computing and functions optimally when it has at least tens of thousands of simultaneous threads. These differences between the CPU and GPU paradigms lead to sharply different programming models, and significant potential performance gains in applications that are able to take advantage of the parallelism a GPU offers. We now provide a brief overview of GPUs—for a more in-depth discussion, see Sanders and Kandrot (2010).

GPUs have been employed in a variety of high-performance computing tasks. While originally designed for graphics, they have also been applied to general-purpose computing (GPGPU) and have been used in a variety of areas—Sanders and Kandrot (2010) list use cases in physics, signal and image processing, computational fluid dynamics, medical imaging, environmental science, deep learning, and a variety of other areas.

There are three major general-purpose programming frameworks for GPUs, *CUDA* (Nickolls et al. 2008), *OpenCL* (Stone et al. 2010), and *HIP* (Zhan and Kindratenko 2016). CUDA is a proprietary framework for Nvidia GPUs, whereas OpenCL is an open standard framework for Nvidia and AMD GPUs, as well as multi-core CPUs, digital signal processors (DSPs), field-programmable gate arrays (FPGAs), and other processors or hardware accelerators. HIP is a CUDA-like interface for Nvidia and AMD GPUs—it was released recently (2016) and is still under heavy development. The frameworks are similar in principle, but scientific computing at present generally focuses on CUDA, because it has a larger user base and better availability of software libraries for primitive tasks such as matrix operations and random number generation. CUDA also supports mature profiling and debugging tools that are not yet widely available in other frameworks. We focus here on Nvidia GPUs and describe them in Nvidia-specific terminology—the structure and terminology for AMD GPUs are largely analogous.

A GPU is organized hierarchically in both hardware and software. The smallest unit that performs computations is a *thread processor* (or *core*). Each thread has access to its index and typically uses it to find its unique data point on which to perform operations. Thread processors are grouped together into *streaming multiprocessors*, which execute code organized into *thread blocks*. A thread block executes groups of 32 threads, called *warps*, in parallel. Each streaming mul-tiprocessor contains *shared memory* that may be accessed by different warps inside the corresponding thread block. The GPU itself is composed of multiple streaming multiprocessors, which have access to *global memory*. A function in CUDA that runs directly on the GPU is called a *kernel* and is launched with respect to a *grid* of thread blocks.

This architecture has a number of implications for performance. Each thread in a warp must perform the same instruction at every point in time. This means that within a warp, if an *if/else* statement is encountered, all threads will first execute the *if* portion, *wait until it completes*, and then execute the *else* portion. This is referred to as *warp divergence* and should be minimized to maintain good performance. Global memory is much slower than other types of memory and should be accessed sparingly—ideally, only to obtain data at the beginning of a computation and to output the results at the end. As noted above, due to both the large number of parallel threads and relatively high latency—which can be hidden by scheduling multiple computations—a GPU should in most cases perform at least tens of thousands of tasks in parallel to be effectively utilized. Though this requirement is not universal—it is easy to construct examples where a smaller number of threads perform better due to memory considerations—it is a good rule of thumb from which to begin.

Thus, for optimal performance, the computation of interest should permit a high degree of what may be termed *fine-grained parallelism*. As previously mentioned, data should be moved to and from the GPU as little as possible—ideally, only at the beginning at end of the computation. Threads should execute code with as few *if/else* statements that evaluate to multiple values within one warp as possible. They should also rely on computations performed by other threads as little as possible, and only by utilizing shared memory within their respective thread block.

Many algorithms will not satisfy these requirements without significant work. For example, difficulties (Lee et al. 2010) can arise in sequential Monte Carlo methods that require frequent bootstrap resampling steps. The most obvious way to parallelize these is via the Poisson bootstrap (Chamandy et al. 2015), which requires drawing Poisson random variables with different parameters in each thread at each iteration. Unfortunately, this presents two problems. Firstly, the Poisson implementation in *cuRAND* requires preprocessing on the CPU for each unique parameter value—a non-starter if our goal is high performance—and hence an alternative GPU-friendly Poisson generation routine would be required. Secondly, the Poisson bootstrap produces a random number of samples at each iteration, potentially creating situations in which too few or too many samples are generated—especially so if particle degeneracy issues are a concern. There are possible alternatives—Murray et al. (2016) have proposed altering the SMC scheme to avoid collective operations. Other difficulties can be present in other algorithms. It is not immediately clear how to parallelize the No-U-Turn sampler (Hoffman and Gelman 2014), a widely used extension of Hamiltonian Monte Carlo which requires building and traversing a binary tree, in non-trivial situations where gradient evaluations alone do not expose sufficient parallelism to saturate the GPU. These algorithms are likely amenable to GPU parallelism, but would require non-trivial modifications to fit the requirements of the GPU's architecture.

We demonstrate below that the class of Gibbs samplers we consider effectively meets all of the desirable criteria mentioned above.

The CUDA Toolkit includes a number of GPU-accelerated libraries. We use *cuBLAS* (for performing matrix operations), *cuRAND* (for parallel random number generation), and *cuSOLVER* (for solving linear systems). Our implementation is written in Scala, a general-purpose compiled language similar to and interoperable with Java. We interface with CUDA using *JCuda*, a set of Java bindings for CUDA, to call GPU code. Custom CUDA kernels are written in CUDA C. Our code is available at https://github.com/aterenin/GPUHorseshoe.

## 4 Exchangeable models

As noted above, to run efficiently on a GPU an algorithm must demonstrate a large amount of parallelism, with minimal interaction between its parts. We now show that data-augmentation-based Gibbs samplers arising from Bayesian statistical models with *exchangeable* likelihoods possess sufficient parallelism.

With vector $\boldsymbol{y} = (y_1, .., y_N)$ of $N \geq 2$ outcome values (in which each $y_i \in \mathbb{R}$), let

$$f(\boldsymbol{y} \mid \boldsymbol{\theta}) = \prod_{i=1}^{N} f(y_i \mid \boldsymbol{\theta}) \tag{1}$$

be an exchangeable likelihood arising from an application of de Finetti's Theorem (de Finetti 1937), and let $\pi(\boldsymbol{\theta})$ be the prior for $\boldsymbol{\theta}$. Suppose that for each $y_i$ there exists a (latent) $z_i$ such that

$$f(y_i \mid \boldsymbol{\theta}) = \int_{\mathbb{R}} f(y_i \mid z_i) f(z_i \mid \boldsymbol{\theta}) \, \mathrm{d}z_i, \tag{2}$$

yielding the data-augmented exchangeable likelihood, which we assume to exist:

$$f(\boldsymbol{y}, \boldsymbol{z} \mid \boldsymbol{\theta}) = \prod_{i=1}^{N} f(y_i \mid z_i) f(z_i \mid \boldsymbol{\theta}). \tag{3}$$

Such likelihoods arise in a large class of models. In fact, complex hierarchical Bayesian models are often built by starting with a set of exchangeability assumptions about the data and specifying a generative process that yields a likelihood expressed in terms of latent variables as above by construction. Now consider the posterior distribution

$$f(\theta, z \mid y) \propto f(y, z \mid \theta)\pi(\theta) \qquad (4)$$

$$= \left[\prod_{i=1}^{N} f(y_i \mid z_i) f(z_i \mid \theta)\right] \pi(\theta). \qquad (5)$$

The full conditional distributions for

$$f(z_i \mid y_i, \theta) \propto f(y_i \mid z_i) f(z_i \mid \theta) \qquad (6)$$

with $i = 1, .., N$, and

$$f(\theta \mid z) \propto \left[\prod_{i=1}^{N} f(z_i \mid \theta)\right] \pi(\theta) \qquad (7)$$

can be used to construct a suitable MCMC algorithm. Often, these full conditional distributions can be recognized from known families of distributions and used to construct a *Gibbs sampler* (Geman and Geman 1984). In what follows, " $\mid -$ " stands for "given all other components of the model."

Algorithm 1.

(a) Initialize arbitrary $(\theta^0, z^0)$.
(b) For $k = 1$ to $k = N_{\text{MC}}$:

   (i) Sample $\theta^k \mid -$.
   (ii) Sample $z_1^k \mid -$.
   (iii) …
   (iv) Sample $z_N^k \mid -$.

where this is not possible, Metropolis–Hastings (Metropolis et al. 1953) steps can instead be performed, yielding a *Metropolis-within-Gibbs* sampler. Both algorithms will in the limit yield samples from the posterior distribution $f(\theta, z \mid y)$, which yields the target distribution $f(\theta \mid y)$ marginally.

Now, suppose that $N$ is large. On a CPU, these algorithms will be slow: there are $N$ full conditionals that each need to be updated at every step of the algorithm. However, notice that $f(\theta \mid z)$ does not depend on the data and that

$$z_i \mid y_i, \theta \perp\!\!\!\perp z_{-i} \mid y_i, \theta \qquad (8)$$

for all $i$ (in which $z_{-i}$ means all of the $z$ values except $z_i$), i.e., all of the variables $z_i$ are full-conditionally independent.

Hence, they can be sampled all at once in parallel on a GPU. This observation forms the basis of our method.

Algorithm 2.

(a) Initialize arbitrary $(\theta^0, z^0)$.
(b) For $k = 1$ to $k = N_{\text{MC}}$:

   (i) Sample $\theta^k \mid -$.
   (ii) Sample $z_1^k \mid -, ..., z_N^k \mid -$ simultaneously in parallel.

Algorithm 2 is a standard MCMC algorithm, so no additional convergence theory for it is needed for it to be valid. Our conditional independence observation is not new—indeed, it is rather obvious and has been known for decades. However, recent advances in GPU programming, such as implementation and wide availability of linear solvers needed for models that include matrix inversion, have made running GPU-accelerated Gibbs sampling tractable, giving a newly available way to counteract issues with mixing and/or slow runtime, provided they are not too severe in the given model. We focus our attention in what follows on implementation and performance.

## 5 Case study: Horseshoe Probit regression

We illustrate GPU-accelerated Gibbs sampling with the Horseshoe Probit regression model (Albert and Chib 1993; Carvalho et al. 2010), which has a likelihood function specified (for $i = 1, \ldots, N$) by

$$y_i \mid z_i = \text{round}[\Phi(z_i)] \qquad z_i \mid x_i, \beta \sim N(x_i\beta, 1) \qquad (9)$$

with $z_i \perp\!\!\!\perp z_{-i}$ given $\beta$, and the following prior:

$$\beta_i \mid \lambda_i, \tau \sim N(0, \lambda_i^2\tau^2) \quad \lambda_i \sim C^+(0, 1) \quad \tau \sim C^+(0, 1). \qquad (10)$$

Here $C^+(0, 1)$ is a standard Cauchy distribution truncated to $\mathbb{R}^+$, $x_i$ and $\beta$ are $(1 \times p)$ and $(p \times 1)$ vectors of data (fixed constants) and regression parameters, respectively, and $y_i \in \{0, 1\}$. The hierarchical nature of the likelihood explicitly specifies the data augmentation in Eq. (3).

We base our algorithm on the Gibbs sampler for probit regression in Albert and Chib (1993), combining it with the hierarchical representation of the Horseshoe and corresponding Gibbs steps in Makalic and Schmidt (2015). By recognizing that for $\theta \in \mathbb{R}^+$,

$$\kappa \sim \text{IG}(1, \theta) \implies \kappa^{-1} \sim \text{G}(1, \theta) = \text{Exp}(\theta), \qquad (11)$$

we arrive at the following full conditionals, suppressing the conditioning bar | henceforth for readability:

$$z_i \sim \text{TN}(\boldsymbol{x}_i \boldsymbol{\beta}, 1, y_i)$$
$$\lambda_j^{-2} \sim \text{Exp}\left[\nu_j^{-1} + \frac{\tau^{-2}\beta_j^2}{2}\right]$$
$$\nu_j^{-1} \sim \text{Exp}(1 + \lambda_j^{-2})$$
$$\boldsymbol{\beta} \sim \text{N}(\boldsymbol{\Sigma}\mathbf{X}^T \boldsymbol{z}, \boldsymbol{\Sigma})$$
$$\tau^{-2} \sim \text{IG}\left[\frac{p+1}{2}, \xi^{-1} + \frac{1}{2}\sum_{j=1}^{p}\lambda_j^{-2}\beta_j^2\right]$$
$$\xi^{-1} \sim \text{Exp}(1 + \tau^{-2}), \qquad (12)$$

where $\mathbf{X}$ is the $(n \times p)$ matrix whose rows are the vectors $\boldsymbol{x}_i$, $\boldsymbol{\Sigma} = (\mathbf{X}^T\mathbf{X} + \tau^{-2}\boldsymbol{\Lambda}^{-2})^{-1}$, $\boldsymbol{\Lambda}^{-2} = \text{diag}(\lambda_1^{-2}, .., \lambda_p^{-2})$, and $\text{TN}(\mu, \sigma^2, y_i)$ is a truncated normal with location $\mu$ and squared scale $\sigma^2$, truncated to $\mathbb{R}^+$ if $y_i = 1$ and $\mathbb{R}^-$ if $y_i = 0$.

Our algorithm performs the following updates:

(i) Sample $\boldsymbol{z}$ and $\boldsymbol{\lambda}^{-2}$ in parallel, followed by $\tau^{-2}$.
(ii) Sample $\boldsymbol{\beta}$, $\boldsymbol{\nu}^{-1}$, and $\xi^{-1}$ in parallel.

We now describe the large degree of parallelism exploited by this sampler and show that every calculation here is well suited to a GPU, in the sense that everything can be performed in parallel with minimal warp divergence.

$\boldsymbol{\beta}$:

i. Precompute $\mathbf{X}^T\mathbf{X}$ before starting the algorithm using *cublasSgemm*.
ii. Calculate $\boldsymbol{\Sigma}^{-1} = \mathbf{X}^T\mathbf{X} + \tau^{-2}\boldsymbol{\Lambda}^{-2}$ using *cublasSaxpy*.
iii. Perform a Cholesky decomposition to find $\mathbf{R}^{-1} = \text{Chol}(\boldsymbol{\Sigma}^{-1})$ using *cusolverDnSpotrf*.
iv. Draw $\boldsymbol{s}$, a vector of IID standard normals, using *curandGenerateNormal*.
v. Compute $\mathbf{R}\boldsymbol{s}$ by solving the triangular system $\mathbf{R}^{-1}\boldsymbol{v} = \boldsymbol{s}$ for $\boldsymbol{v}$ using *cublasStrsv*.
vi. Compute $\mathbf{X}^T \boldsymbol{z}$ using *cublasSgemv*.
vii. Compute $\boldsymbol{\mu} = \boldsymbol{\Sigma}\mathbf{X}^T \boldsymbol{z}$ by solving the linear system $\boldsymbol{\Sigma}^{-1}\boldsymbol{\mu} = \mathbf{X}^T \boldsymbol{z}$ for $\boldsymbol{\mu}$ (reusing the already-computed Cholesky decomposition), using *cusolverDnSpotrs*.
viii. Compute $\boldsymbol{\beta} = \mathbf{R}\boldsymbol{s} + \boldsymbol{\mu}$ using *cublasSaxpy*.

$\boldsymbol{z}$:

i. Calculate $\boldsymbol{\mu} = \mathbf{X}\boldsymbol{\beta}$ using *cublasSgemv*.
ii. Draw $\boldsymbol{z}$ by using the following sampling routine, implemented as a custom CUDA kernel.

   a. Compute $y_i^* = 2y_i - 1$ and $\mu_i^* = \mu_i y_i^*$. Note that if $y_i = 1$ then $y_i^* = 1$, and if $y_i = 0$ then $y_i^* = -1$.
   b. If $\mu^* > 0.47$, use a Gaussian rejection sampler.
      i. Draw $s_i \sim \text{N}(0, 1)$ and compute $z_i^{\text{prop}} = s_i + \mu_i^*$.
      ii. Return $z_i = y_i^* z_i^{\text{prop}}$ if $z_i^{\text{prop}} > 0$. Otherwise, draw a new Gaussian and try again.
   c. Else, use the Exponential rejection sampler described in Robert (1995).
      i. Calculate the proposal parameter $\theta_i$ in Robert (1995). The formula for $\theta_i$ is modified for nonzero mean and truncation to $\mathbb{R}^+$, and the acceptance probability $\varrho_i$ is modified accordingly as well.
      ii. Draw $u_i^{(1)} \sim \text{U}(0, 1)$ and (independently) $u_i^{(2)} \sim \text{U}(0, 1)$, compute $z_i^{\text{prop}} = -\theta_i^{-1} \ln(u_i^{(1)}) \sim \text{Exp}(\theta_i)$, and $\varrho_i$.
      iii. Return $z_i = y_i^* z_i^{\text{prop}}$ if $u_i^{(2)} < \varrho_i$. Otherwise, draw two new independent uniforms and try again.

Note that there are three possible kinds of warp divergence in this routine that need to be minimized. The first kind arises because the two rejection samplers cannot be performed in parallel—this is unavoidable unless all threads in a warp have $\mu^* < 0.47$ or $\mu^* \geq 0.47$. The second kind arises from the iterative nature of rejection sampling: a warp will finish only when *all* threads have accepted—this is kept under control because both samplers are efficient, with an overall worst-case acceptance probability of around 2/3 for $\mu^*$ in a neighborhood around 0.47, and probability near 1 everywhere else. The third kind arises because some $z_i$ will be truncated to $\mathbb{R}^+$ while others are truncated to $\mathbb{R}^-$—this is eliminated completely by introducing the $y_i^*$.

$\boldsymbol{\lambda}^{-2}, \boldsymbol{\nu}^{-1}, \xi$:

i. Draw $\boldsymbol{u}$, a vector of IID uniforms, using *curandGenerateUniform*.
ii. Calculate the parameter vector $\boldsymbol{\theta}$ of the exponential distributions in question, and transform $\boldsymbol{u}$ to $\text{Exp}(\boldsymbol{\theta})$ via the inversion formula $-\boldsymbol{\theta}^{-1} \ln(\boldsymbol{u})$, implemented as a custom CUDA kernel.

$\tau^{-2}$:

i. Calculate $\boldsymbol{\beta}^2$ using a custom CUDA kernel.
ii. Calculate the sum in the scale parameter by expressing it as the dot product $\boldsymbol{\lambda}^{-2} \cdot \boldsymbol{\beta}^2$ using *cublasSdot*.
iii. Draw $\tau^{-2} \sim \text{G}(a, b)$ using the rejection sampler in Cheng (1977), implemented as a custom CUDA kernel. This kernel involves cooperation among threads and is described as follows.

   a. Each of the 32 threads draws two uniforms, performs an accept–reject step independently, and writes

    its results to shared memory within the block, after which all threads synchronize.
  b. Thread 1 sequentially looks through shared memory and returns once it finds an accepted value.
  c. If all threads reject, new proposals are computed. This is exceedingly rare for $b$ large: each thread accepts with probability approximately 0.88 (Cheng 1977), so the combined acceptance probability for the entire kernel is about $1 - 3.4 \times 10^{-30}$.

Our current implementation is well optimized in some ways and suboptimal in others. The update for $\beta$ given the precision matrix avoids matrix inversion in favor of solving a linear system and a triangular system via a single Cholesky decomposition. This works because $\Sigma = \mathbf{R}^T \mathbf{R} \implies \Sigma^{-1} = \mathbf{R}^{-1}(\mathbf{R}^{-1})^T$. This is far more efficient than the following process, which we call the naive implementation because it matches up well with mathematical formulas but is a poor choice for performance:

(1) calculate and store $(\mathbf{X}^T\mathbf{X} + \Sigma_0)^{-1}$ by behind-the-scenes performing an LU decomposition,
(2) calculate $(\mathbf{X}^T\mathbf{X} + \Sigma_0)^{-1}\mathbf{X}^T z$,
(3) use the *R* function *mvrnorm* in the package *MASS* (Venables and Ripley 2002) to sample from a multivariate Gaussian by behind-the-scenes performing an eigenvalue–eigenvector decomposition of $(\mathbf{X^T X} + \Sigma_0)^{-1}$.

In addition to extra matrix decompositions that degrade performance and numerical stability, the naive implementation also calculates and stores a matrix inverse, which in general should never be done when that matrix inverse will subsequently be multiplied by another matrix or vector, as solving a linear system is mathematically equivalent and computationally cheaper. Our routine performs substantially better than the naive approach, from both a speed and stability standpoint, though our approach is also not perfect from a linear algebra standpoint. It would be more numerically stable to instead use a *QR* decomposition with rank-1 updates in place of the Cholesky, as this would avoid the construction of $\mathbf{X}^T\mathbf{X}$—we did not pursue this route because no such routine is implemented in *cuSOLVER*. An even more stable, albeit slower, approach would involve replacing the *QR* decomposition with *SVD*.

Most of the steps in the above routines parallelize immediately—the *cuBLAS*, *cuRAND*, and *cuSOLVER* library calls are highly optimized to the GPU. Our custom CUDA kernels, on the other hand, are not. The update for $z$ could be done more efficiently by, once after burn-in, sorting the data according to the mean vector $\mathbf{X}\beta$, which would minimize warp divergence as it would ensure that most warps perform the same kind of rejection sampler. Due to the additional code required to track the order of the sorted data on the GPU, we did not pursue this idea.

The update for $\tau^{-2}$ could be made more efficient through the use of atomic operations or shuffle instructions in the final step. We did not pursue these because $\tau^{-2}$ is updated in parallel with $z$, and its computational burden is comparatively tiny with little impact on performance. We note, however, that computing $\tau^{-2}$ would likely not be faster if executed by fewer than 32 threads, as the GPU would need to use a 32-thread warp to run the CUDA kernel even if some threads are doing nothing.

Our implementation mixes the *cuRAND host API*, which generates fixed-size arrays of random numbers in parallel, and the *cuRAND device API*, which allows each thread to generate its own random numbers. We use the host API as much as possible, because it is far easier to implement and manage. The device API requires management of random number generator seeds and offsets on a per-thread basis. The simplest approach is to have separate random number generators in each thread, seeded to ensure that random numbers do not overlap. The main drawback of this approach is that it requires each thread to store its random number generator state in global memory. Instead, our code stores only one random generator state and works by

(1) copying the random number generator state from global memory to local memory,
(2) skipping forward $k$ iterations in the random number generator, to ensure that random numbers generated by different threads do not overlap,
(3) generating random numbers, and
(4) once the computation is complete, writing the last thread's random number generator state back to global memory.

To be efficient, this requires a random number generator capable of skipping ahead an arbitrarily large number of iterations in constant time without causing warp divergence. The *cuRAND* default *XORWOW* random number generator (Marsaglia 2003) does not have constant-time skip ahead, and neither does the parallel version of the popular *Mersenne Twister* (Matsumoto and Nishimura 1998), but the counter-based *Philox* random number generator of Salmon et al. (2011) does. We highly recommend *Philox* for its parallel efficiency, and because it is *Crush-resistant* (i.e., it passes all of the tests in *TestU01* (L'Ecuyer and Simard 2007), a suite for examining the quality of random number generators)—see Manssen et al. (2012) for a review of parallel random number generators. *Philox* is especially well suited to situations, such as ours, where random numbers cannot be pre-generated because the amount of random numbers needed per-thread changes at every iteration.

To further accelerate the algorithm, we overlap computation and data transfer (IO) steps—an example of *hiding latency*. For example, once $\boldsymbol{\beta}$ has been successfully updated, Monte Carlo output is downloaded off of the GPU at the same time as updates for other variables are computed. This is done by utilizing CUDA *streams*, a programmatic concept in CUDA that allows multiple kernels and memory operations to execute simultaneously in parallel.

We ran our algorithm on the Horseshoe Probit regression model (9, 10) with multiple data sets, varying $(N, p)$. Performance is summarized in the following section.

## 6 Performance results

### 6.1 Synthetic data

We first ran our algorithm on a synthetic data set with known correct answer, generated by taking

$$x_{ij} \sim N(0, 1)$$
$$\boldsymbol{\beta} = (1.3, 4, -1, 1.6, 5, -2, \mathbf{0}_{p-6})^T$$
$$\boldsymbol{y} \sim \text{Ber}\left[\Phi(\mathbf{X}\boldsymbol{\beta})\right]. \tag{13}$$

Our goal was to study the GPU implementation on a nontrivial model—see Chopin and Ridgway (2015) for a recent discussion on the use of binary regression models for comparing performance of Bayesian algorithms. The GPU used was a Nvidia GeForce Titan X with about 12 GB RAM and single-precision floating-point performance of about 11 TFLOPs.

To compare CPU and GPU run times, we ran the algorithm on both a workstation and a laptop. The laptop ran a single-threaded *R* implementation on an Intel Core i5 3210M CPU, and the workstation ran a multi-threaded *Scala* implementation on two Intel Xeon E5-2650 CPUs for a total of 16 cores. On the laptop, we used $N$ up to 1000 and $p$ up to 100. On the workstation, we used $N$ up to 1,000,000 and $p$ up to 1000. Both the single-threaded *R* and multi-threaded *Scala* implementations used the same optimized precision-matrix-based sampling scheme for $\boldsymbol{\beta}$ as in Sect. 5 and likewise for all other variables. In particular, the *Scala* implementation was parallelized in the same way as the GPU implementation—using exchangeability together with multi-threaded BLAS/LAPACK calls and random number generation. The resulting comparison showcases expected performance for interpreted and compiled languages, respectively. This provides a reasonable comparison with performance that users might expect on their laptops and on remotely accessible workstations.

Then, to study scaling, we selected a variety of different combinations $(N, p)$ with $N$ up to 1,000,000 and $p$ up to 10,000. Due to memory limitations, we did not run $N = 1,000,000$ and $p = 10,000$ simultaneously, as this would have required over 40 GB RAM. Larger data sets are possible but more difficult to accommodate, because such data sets would require either multiple GPUs or streaming implementations, and we chose not to explore that possibility here—see Sect. 7.

We ran the Gibbs sampler for 10,000 iterations in all cases, starting from $z$, $\boldsymbol{\beta}$ set to 0 and $\boldsymbol{\lambda}, \boldsymbol{\nu}, \tau, \xi$ set to 1. The Horseshoe posterior correctly identified all nonzero values in $\boldsymbol{\beta}$ and shrank irrelevant coefficients to zero. Mixing leaves much to be desired for large $N$ and $p$, but is data-dependent, so we focus on mixing in the context of real data—see Sect. 6.2.

Run time is shown in Fig. 1, which shows how different values of $N$ and $p$ affect GPU run time, along with the 16-threaded dual-CPU workstation run time in *Scala*, and single-threaded laptop CPU run time in *R* for comparison purposes. All times referred to in this paper are clock time.

The GPU is clearly many times faster than the laptop for all data sizes we examined—for instance, with $(N, p) =$

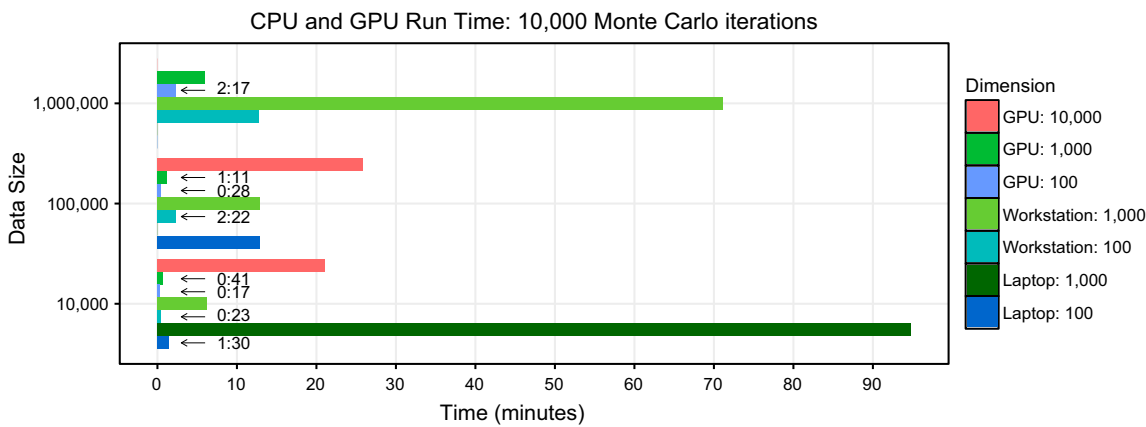

**Fig. 1** Comparison of GPU, workstation (dual 8-core CPUs), and laptop (single-core CPU) run time for 10,000 iterations of the Horseshoe Probit Gibbs sampler with synthetic data

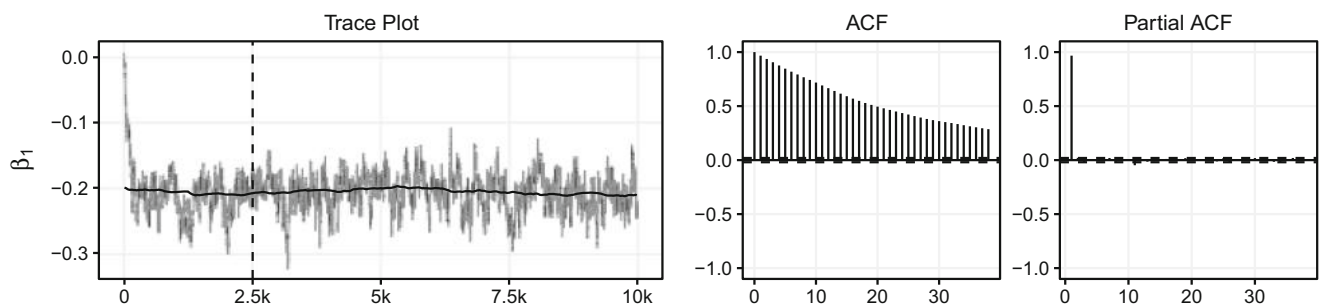

**Fig. 2** Diagnostics for $\beta_1$ in the real data example of Sect. 6.2, based on 10,000 Monte Carlo samples

(10,000, 1000), the CPU and GPU calculations took 95 min and 41 s, respectively. The GPU is also significantly faster than the more expensive dual-CPU workstation for sufficiently large problems: for $(N, p) = (1,000,000, 1000)$, the CPU and GPU calculations took 71 and 2 min, respectively. Furthermore, a tenfold increase in data set size does not necessarily result in a tenfold increase in GPU run time. This is because larger data set sizes expose a larger degree of parallelism that the GPU is able to take advantage of. This is evident for $N = 10,000$ going to $N = 100,000$, and for $N = 100,000$ going to $N = 1,000,000$—in both cases, a tenfold increase in data set size increases run time on the GPU by a factor ranging from 1.4 to 2.1. The GPU is faster than two 8-core CPUs even when the data size is increased tenfold: going from $(N, p) = (100,000, 1000)$ on the workstation to $(1,000,000, 1000)$ on GPU *reduces* run time by a factor of 2.2. By profiling our program, we found that it spent the vast majority of time performing matrix-vector multiplications, indicating that there is little we could have done to further speed up our implementation.

## 6.2 Real data

One of us (DD) was recently involved in a study in which the goal was to predict, at time $t$ for patients hospitalized in a general medical ward, whether or not such patients would have an unplanned transfer to the intensive care unit (an adverse outcome to be avoided if possible) in a 4-h time window starting at time $(t + 8)$ h. Separate models were built for four sets of principal diagnoses at admission—gastrointestinal bleeding, serious heart problems (heart attack, congestive heart failure), pneumonia and sepsis (a total-body infection)—plus a category consisting of all other major admission diagnoses that often lead to unplanned transfers. Focusing on patients in the admission category Other, we created a data set with $N = 1,295,848$ 1-h hospitalization episodes, which were rendered conditionally exchangeable by a rich set of 211 clinical predictor variables. At the 1-h episode level, the adverse outcome was rare (our dichotomous outcome variable had a mean of 0.0048) but highly relevant to the appropriate care path, so accurate prediction was clinically crucial.

Preliminary descriptive analysis narrowed the available independent variables down to a set of $p = 141$ interesting predictors—where *interesting* was determined according to signal-to-noise ratios in maximum-likelihood estimation—and we then used our algorithm to fit a Horseshoe Probit regression model to the resulting data set, to see how many of the interesting predictors survived the regularization process imposed by the Horseshoe prior.

From a cold start, in which the $z$ and $\beta$ were all initialized to 0 and $\lambda, \nu, \tau, \xi$ were started at 1, our GPU algorithm produced 10,000 iterations—2500 for burn-in and 7500 for monitoring—in 2.94 min. All of the components of the $\beta$ vector reached equilibrium from the cold start in about 100–200 iterations. Their mixing was slow—a typical $\beta_i$ had an output trace that behaved like an $AR_1(\rho)$ time series with first-order autocorrelation $0.88 < \rho < 0.98$. Trace plots are shown in Fig. 2. To ensure reliable inference, we re-ran the algorithm for 500,000 iterations, which yielded a near-identical distribution.

The Horseshoe prior shrunk about 20% of the initially interesting predictor coefficients sharply back toward 0 (using 90% posterior intervals), leading to a more parsimonious model with a perceptible improvement in predictive performance.

## 7 Discussion

Even though MCMC methods such as Gibbs sampling are inherently sequential, our results demonstrate that the steps needed at each iteration can be sufficiently parallelizable that Gibbs sampling may be made to run orders of magnitude faster on a GPU than a CPU. We have found this to be true for our model, and we expect it to be true for other models. Exchangeable models look particularly promising, because they yield full-conditionally independent latent variables, which can immediately be updated as a block on a GPU. Models with more complicated forms of exchangeability parallelize in more complicated ways—for an example involving Latent Dirichlet allocation (Blei et al. 2003), see Magnusson et al. (2015) and Terenin et al. (2017). Many non-exchangeable models, such as Gaussian Processes and Dirichlet Process Mixtures, can likely also benefit from GPU acceleration, if there exists a sufficient degree of available parallelism in the Gibbs steps. In our implementation, updating the $\boldsymbol{\beta}$ vector in the Horseshoe Probit regression model is significantly faster on a GPU, in spite of the fact that $\beta_i \not\perp\!\!\!\perp \beta_j$ for $i \neq j$.

The parallelism inherent in exchangeability is present other algorithms. For instance, we could have used Hamiltonian Monte Carlo for $\boldsymbol{\beta}$ in our Horseshoe Probit model instead of the data augmentation of Albert and Chib (1993), eliminating $z$ but retaining all other parameters. This *HMC-within-Gibbs* sampler would substantially improve mixing, and gradient evaluations for $\boldsymbol{\beta}$ could be performed in parallel—see Beam et al. (2015). Note, however, that all other steps would be identical to the ones described here. We have thus chosen to focus on Gibbs sampling because of its use as a basic building block in more complicated algorithms.

The larger versions of our synthetic data sets were large enough to take up most of the GPU's memory, so $(N, p) = (1{,}000{,}000, 1000)$ was the biggest problem size we explored. There are several approaches to computing with data sets that are too big to fit in a GPU's memory. Multiple GPUs can be used: this will introduce some performance penalty due to synchronization requirements, but may work quite well in problems that are largely limited by computationally intensive data-parallel steps. Bigger data sets may also be analyzed by using clusters of GPUs with Asynchronous Gibbs sampling—see Terenin et al. (2016). Alternatively, data can be streamed through the GPU as it is needed, rather than loaded into memory before beginning the computation. This approach has been used for point estimation in large-scale models in software packages such as *Vowpal Wabbit* (Langford 2007) on CPUs and *Bidmach* (Canny and Zhao 2013) on GPUs and in our view would likely also work well for Gibbs sampling.

The Nvidia GeForce Titan X GPU that we employed is a high-end consumer GPU designed for video gaming and lightweight scientific computing use cases. It is well suited to MCMC methods, because they are robust to numerical precision and can successfully run in floating point—see Breyer et al. (2001) and Mingas and Bouganis (2012).

GPUs are becoming increasingly accessible. When this was written, servers with scientific-computing-grade GPUs can be rented, from a variety of cloud computing providers, for well under 1 US dollar per hour. With the recent release of the Thunderbolt 3 specification, it is also possible, through the use of an external GPU case such as the Akitio Node, to connect a desktop-grade GPU to a laptop computer using a simple cable.

Software is currently the most significant barrier to widespread adoption of GPU acceleration for Bayesian inference. The functions provided by CUDA libraries are low level and present a high degree of difficulty to those not already familiar with high-performance computing libraries such as BLAS and LAPACK. New frameworks are needed to bring modern programmatic concepts into the GPU software stack. These would parallel recent advances in programmatic concepts and frameworks for compute clusters, such as *Akka* (Typesafe Inc. 2015) and *Apache Spark* (Zaharia et al. 2010), which made computation in the cluster environment much more user-friendly through the implementation of modern concepts such as actor models, functional programming, and lazy evaluation. Similar work is needed in simplifying GPU debugging.

This work has already begun: recent computational graph frameworks such as TensorFlow (Abadi et al. 2017) are allowing GPU code to be generated automatically for many classes of problems—in our view, second-generation frameworks, such as DyNet (Neubig et al. 2017) and PyTorch (Paszke et al. 2017), are especially promising due to their increasing flexibility and ease of use. We hope that these frameworks begin to find adoption outside of the deep learning community in which they were originally proposed and that perhaps in a few years it will be just as easy to write a Gibbs sampler on a GPU as it currently is on a CPU in a high-level language such as *R*.

Another difficulty in CUDA is lack of random number generation routines for less-common distributions. In the Horseshoe Probit regression model, we needed to implement custom CUDA kernels for sampling Exponential, Gamma, and Truncated Normal random variables. We also needed to implement a CUDA kernel for squaring arrays of numbers—while trivial to write, this resulted in yet another block of code that needed to be written, debugged, and compiled before we could run our algorithm. These issues are even more pronounced in OpenCL, where there are no widely available libraries for many matrix and vector routines. Future work will likely remove these difficulties.

Though our initial results on the Horseshoe Probit regression model are promising, further work is needed to study Bayesian computation on GPUs. Most parallelization strategies are unexplored, and the relative performance of different methods is unknown. We believe that our work in this case study demonstrates that fast generic methods for large classes of Bayesian inferential and predictive problems are possible and that GPU-accelerated Bayesian computation is worth substantial further exploration.

**Acknowledgements** The authors gratefully acknowledge Nvidia and Akitio Inc. for providing the hardware used in this research. S.D. was partially supported by NSF Grant OAC-1541270.